\documentclass[aps,prl,twocolumn,showpacs]{revtex4}
\usepackage{graphicx}

\begin{document}

\author{S. O'Brien$^{1}$, S. Osborne$^{1}$, K. Buckley$^{1}$, R. Fehse$^{1}$, A. Amann$^{1}$, E. P. O'Reilly$^{1}$, 
L. P. Barry$^{2}$, P. Anandarajah$^{2}$, J. Patchell$^{3}$ and J. O'Gorman$^{3}$}

\affiliation{$^{1}$Tyndall National Institute, University College, Lee Maltings, Cork, Ireland \\
$^{2}$RINCE, School of Electronic Engineering, Dublin City University, Dublin 9, Ireland \\
$^{3}$Eblana Photonics, Trinity College Enterprise Centre, Pearse Street, Dublin 2, Ireland}

\title{Two-Color Fabry-P\'erot Laser Diode with THz Primary Mode Spacing}

\begin{abstract}

A class of multiwavelength Fabry-P\'erot lasers is introduced where the spectrum 
is tailored through a non-periodic patterning of the cavity effective index. The 
cavity geometry is obtained using an inverse scattering approach and can be designed 
such that the spacing of discrete Fabry-P\'erot lasing modes is limited only by the 
bandwidth of the inverted gain medium. A specific two-color semiconductor laser with 
a mode spacing in the THz regime is designed, and measurements are presented 
demonstrating the simultaneous oscillation of the two wavelengths. The extension of 
the Fabry-P\'erot laser concept described presents significant new possibilities in 
laser cavity design. 

\end{abstract}


\maketitle

The most familiar laser cavity geometry is the Fabry-P\'erot (FP) laser, which comprises 
an active gain medium and two external mirrors providing feedback for oscillation. In this 
geometry the longitudinal lasing mode wavelengths are determined by the half wave resonance 
condition: $\lambda^{m} = 2 n L_{c}/m$, where $m = 1,2,...$, $\lambda^{m}$ is the free space 
wavelength of the $m\mbox{th}$ mode, $L_{c}$ is the cavity length,
and $n$ is the cavity refractive index \cite{agrawal}. 

A fundamental limitation of the basic FP geometry is the lack of any frequency selectivity 
other than that provided by the gain medium. Because the gain bandwidth is much larger 
than the FP mode spacing in typical semiconductor lasers, more complex laser cavity geometries 
have been conceived in order to control and manipulate semiconductor laser 
spectra. For example, one dimensional systems such as the distributed feedback laser provide high 
spectral purity and temperature stability in device applications \cite{kogelnik}. Translational 
symmetry determines the lasing modes of this structure without the need for a reflection from 
external mirrors. 

If we consider the interaction of the cavity modes with the gain medium, the 
semiconductor FP laser geometry is deceptively simple. In a perfectly homogeneously broadened 
medium, a single lasing mode should always dominate \cite{haken, tredicucci}. When driven 
above threshold, semiconductor FP lasers often oscillate in many modes. This multimode behaviour 
is characteristic of inhomogeneously broadened gain media, despite the fact that in a 
semiconductor the carriers are distributed in continuous bands. 

A key property of the FP laser in this respect is the fact that all the cavity resonant 
wavevectors are equally spaced. As a result, four-wave mixing (FWM) interactions, which 
transfer power among modes, are cavity enhanced. In addition, carrier density pulsations 
at the intermode frequencies and the finite alpha factor lead to an asymmetric contribution 
to the nonlinear gain in semiconductor lasers \cite{yamada, uskov, ogita}. Along with the 
large spontaneous emission rate in semiconductor lasers, this interaction also promotes 
multimode oscillation and can lead to mode hopping and complex antiphased switching dynamics 
\cite{ahmed, yacomotti}. 

In this letter we revisit and extend the basic FP laser geometry. We demonstrate that 
multiwavelength FP lasers can be designed where, apart from the constraint imposed by the 
half wave resonance requirement, the distribution of lasing modes is chosen 
\textit{independent} of the cavity length. The basic FP cavity configuration and mode 
structure are maintained, with the manipulation of the lasing mode spectrum achieved using 
a non-periodic effective index profile. The precise geometry is determined from the desired 
lasing spectrum through an inverse scattering approach \cite{obrien,obriens_josaB}. 

Because of its fundamental significance, we present experimental measurements of a two-color 
semiconductor FP laser with a primary mode spacing in the THz regime. Our measurements demonstrate 
that the device oscillates simultaneously on two discrete FP modes, and without the requirement for an 
external cavity arrangement or other external perturbation. In contrast to this ideal behaviour associated 
with weakly coupled modes, an otherwise identical \textit{plain} FP laser, with a modal spacing 
determined by the cavity length, displays the mode hopping behaviour and complex dynamics associated 
with strong mode competition. 

\begin{figure}
\includegraphics[height=4.0cm,width=7.5cm]{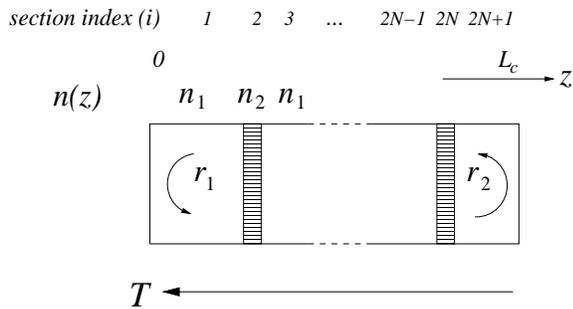}
\caption{\label{capl1} One dimensional model of a Fabry-P\'erot laser cavity 
of length $L_{c}$ and including $N$ index steps. The cavity effective 
index is $n_{1}$ while the additional features providing the index step (shaded regions) 
have effective index $n_{2}$ as shown. All cavity sections are numbered $1 \leq i \leq 2N+1$ 
beginning on the left. The matrix $T$ relates the left and right moving fields 
inside the cavity at the cavity mirrors. The mirror reflectivities are 
$r_{1}$ and $r_{2}$ as shown.}
\end{figure}

Consider the one-dimensional model of the FP cavity geometry represented 
in Fig. \ref{capl1}. The system comprises a FP cavity of length $L_{c}$ with a spatially 
varying refractive index. The mirror reflectivities are $r_{1}$ and $r_{2}$ (assumed real 
for simplicity) and there are $N$ additional index steps along the cavity. For each section 
of the laser cavity (index $i$) we define $\theta_{i} = n_{i}k_{0z}L_{i}$, where $k_{0z}$ is
the free space wavenumber along $z$ and $L_{i}$ and $n_{i}$ are the length and the effective 
refractive index of the $i\mbox{th}$ section respectively. The adjusted complex optical path 
length across the cavity is then $\sum_{i=1}^{2N+1}\theta_{i}$. 

We set the background cavity effective index, $n_{1} = n$, and the effective 
index at the index step features, $n_{2} = n + \Delta n$. Suppose the transfer matrix $T$ relates the 
right and left moving electric fields, $E^{\pm}(z)$, at the cavity mirrors in Fig. \ref{capl1}. Then the 
lasing modes of the cavity are defined by the relation \cite{obriens_josaB}
\begin{equation}\label{threshold}
\left(
\begin{array}{c}
1 \\
r_{1}
\end{array}
\right) = T \left(
\begin{array}{c}
r_{2} \\
1
\end{array}
\right).
\end{equation}
From Eqn. \ref{threshold} one can show that the lasing condition at first order in the index step can 
be written as 
\begin{eqnarray}\label{gt_cond}
&1 - r_{1}r_{2}\exp(2i\Sigma \theta_{i}) =  \nonumber \\
&i\frac{\Delta n}{n} \sum_{j} \sin \theta_{2j}
\left[r_{1}\exp(2i\phi_{j}^{-}) + r_{2}\exp(2i\phi_{j}^{+})\right], 
\end{eqnarray}
where the quantities $\phi^{-}_{j} \mbox{ and } \phi^{+}_{j}$ are the optical path lengths 
from the centre of each additional feature to the left and right facets respectively. 

If we neglect a factor describing the background losses associated with each mode, for a vanishing index step, 
the threshold gain for lasing is determined by the mirror losses. We have $\gamma_{m}^{(0)} = L_{c}^{-1} 
\ln 1/r_{1}r_{2}$. In the perturbed case, a set of self-consistent equations for the lasing modes 
is found by making an expansion in Eqn. \ref{gt_cond} about the cavity resonance condition: 
$\sum \theta_{i}^{'} = {\phi^{-}_{j}}^{'} + {\phi^{+}_{j}}^{'} = m\pi + 
\delta_{m}$ \cite{obrien}, where $\delta_{m} (\ll 1)$ determines the lasing mode frequency shift. 

The inverse problem at first order is solved by choosing a particular cavity resonance, $m_{0}$, as 
an origin in wavenumber space. We assume quarter wave features with $\sin \theta_{2j}^{'} = 1$ 
in order that the intensity scattered by each feature at the wavelength of mode $m_{0}$ is maximized. 
Each feature is placed ultimately such that a half wavelength subcavity at the wavelength of mode 
$m_{0}$ is formed between the feature and one of the external mirrors. The threshold gain can then 
be expressed at each resonance, $m$, where $m = m_{0} + \Delta m$, as
$\gamma_{m} = \gamma_{m}^{(0)} + (\Delta n/n) \gamma_{m}^{(1)}$, where 
\begin{eqnarray}\label{gamma_m1}
&\gamma_{m}^{(1)} = \frac{1}{L_{c}\sqrt{r_{1}r_{2}}}\cos(m_{0}\pi) \cos(\Delta m\pi)  \nonumber \\
& \times \sum_{j=1}^{N}A(\epsilon_{j}) \sin(2\pi\epsilon_{j}m_{0}) \cos(2\pi\epsilon_{j}\Delta m).
\end{eqnarray}
In the above expression, the factor $A(\epsilon_{j}) = r_{1} \exp(\epsilon_{j}L_{c}\gamma_{m}^{(0)}) 
- r_{2} \exp(-\epsilon_{j}L_{c}\gamma_{m}^{(0)})$ and $\epsilon_{j}$ is the position of the center of each 
feature measured from the center of the cavity as a fraction of the cavity length. 

\begin{figure}
\includegraphics[height=4.5cm,width=7.5cm]{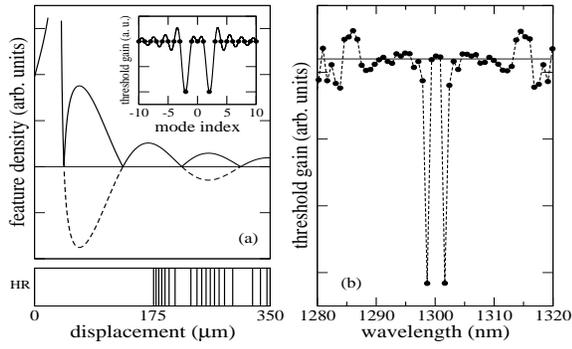}
\caption{\label{ob_ft} (a) Feature density function (solid line). The dashed lines are the negative of 
the feature density function in those intervals where the Fourier transform of the function shown in the inset 
is negative. Inset: Ideal threshold gain of modes in wavenumber space. Lower panel: Laser cavity schematic 
indicating the locations of the additional features. The device is high-reflection (HR) coated as indicated. 
(b) Calculation of the threshold gain of modes for the laser cavity schematically pictured in the lower panel 
of the figure. The horizontal line is at the value of the mirror losses of the plain cavity.}
\end{figure}

Using Eqn. \ref{gamma_m1} Fourier analysis can be used in order to build up a particular threshold gain 
modulation in wavenumber space. Because it represents the simplest system that illustrates the application 
of classical multimode laser theory \cite{sargent}, the  multiwavelength device we describe here is the 
two-color laser. The appropriate basis for this device is a pair of sinc functions, $\gamma_{m} \sim 
\mbox{sinc}(\Delta m + a/2) + \mbox{sinc}(\Delta m - a/2)$. This choice selects two modes, centered at 
$m_{0}$ and with spacing $a$ modes, while leaving the other FP modes unperturbed. In the inset of Fig. 
\ref{ob_ft} (a) we have plotted an idealized threshold gain spectrum where the primary mode spacing is 
$a = 4$ fundamental cavity modes. The Fourier transform of our idealized threshold gain modulation 
function is $\cos(\pi a \epsilon)$, for $-1/2 \leq \epsilon \leq 1/2$ and is zero otherwise. 

The factor $A(\epsilon_{j})$ in Eqn. \ref{gamma_m1} reflects the fact that the change in threshold of a given mode 
is determined by the difference in the round-trip amplitude gain to the left and to the right of each feature. 
To determine the appropriate distribution of features, we must therefore correct for the variation of the 
amplitude of the threshold modulation with position. We take the product of the Fourier transform of our 
ideal threshold gain modulation with the envelope function, $[A(\epsilon_{j})]^{-1}$. The absolute value of 
this product determines the feature density function shown in Fig. \ref{ob_ft} (a), which we then sample in 
order to approximately reproduce the idealized threshold gain spectrum \cite{obrien}. 

For the two-color device considered the cavity is asymmetric with one larger facet reflectivity $(r_{1})$. This allows 
a more uniform density of features along one side of the device center.
Once the feature density function is sampled correctly, feature positions are adjusted in order
to satisfy the correct phase requirement for resonance. A schematic picture of the device, high-reflection  
coated as indicated, is shown in the lower panel of Fig. \ref{ob_ft}. With respect to the lasing wavelength of 
mode $m_{0}$ in the cavity, where $\cos(\pi a \epsilon_{j}) > 0$, the phase requirement corresponds to forming a 
halfwave resonant subcavity between the corresponding feature and the high-reflection coated mirror. 
For $\cos(\pi a \epsilon_{j})
< 0$, we form a quarterwave subcavity at the same wavelength. In this way, at each zero of the feature density 
function a $\pi/2$ phase shift is introduced into the index pattern along the device length. Optical path 
corrections due to the introduction of the features must also be accounted for when the final feature positions are 
calculated. The calculated form of the threshold gain spectrum is shown in Fig. \ref{ob_ft} (b) and is an excellent 
approximation of the ideal form. To appreciate the simplicity of this approach, it is important to note that the device 
comprises a single, patterned amplifying section with both external mirrors necessary to form the FP mode structure. 
Thus, unlike distributed feedback approaches, the role of the inhomogeneous cavity effective index is simply to 
discriminate between the various FP modes. 

We now present experimental measurements of a ridge waveguide FP laser fabricated to the design depicted in Fig. 
\ref{ob_ft}.  The device is a  multi-quantum well InP/InGaAlAs laser of length 350 $\mu$m with a peak emission 
near 1.3$\mu$m.  The additional features are slotted regions etched into the laser ridge waveguide. This technique 
is based on standard optical lithography and does not require a regrowth step. The laser was temperature 
stabilised at 25$^{0}$C to $\pm 0.01^{0}$C and a constant current was applied to the device. Laser emission 
was spectrally resolved using an optical spectrum analyser with 0.01 nm resolution. 

\begin{figure}
\includegraphics[height=5.0cm,width=8.0cm]{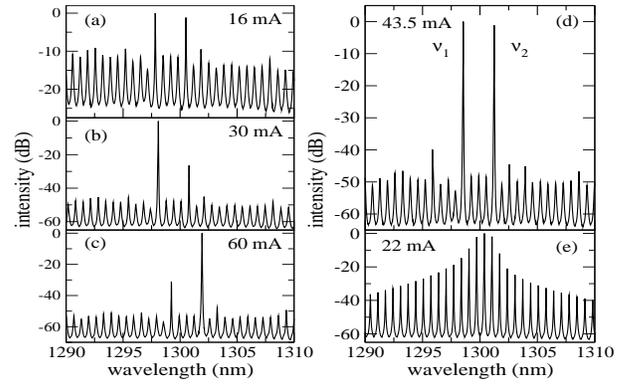}
\caption{\label{spectra} (a) Below threshold spectrum of the two-color device of Fig. \ref{ob_ft}. 
(b) Lasing spectrum at 30 mA. (c) Lasing spectrum at 60 mA. (d) Two-color lasing spectrum at 43.5 mA 
(e) Above threshold spectrum of a plain Fabry-P\'erot device fabricated on the same bar.}
\end{figure}

A series of spectra of the device of Fig. \ref{ob_ft} are shown in Fig. \ref{spectra}. Fig. 
\ref{spectra} (a) shows the device spectrum below threshold. One can see that the two primary modes are 
already selected in this regime. Note the good agreement with the calculation shown in Fig. \ref{ob_ft}, 
with the two primary modes separated by four fundamental FP modes. As the current is increased the mode 
on the short wavelength side reaches threshold first (Fig. \ref{spectra} (b)) and as the current is increased 
further thermal effects lead to the peak power shifting across the primary mode spacing to the long wavelength 
side (Fig. \ref{spectra} (c)). The lasing spectrum at 43.5 mA  is shown in Fig. \ref{spectra} (d). At this 
current the time averaged optical power in the primary modes is approximately equal. For comparison, an above 
threshold spectrum from a plain FP laser fabricated on the same bar is shown in  Fig. \ref{spectra} (e). 

\begin{figure}
\includegraphics[height=4.5cm,width=7.5cm]{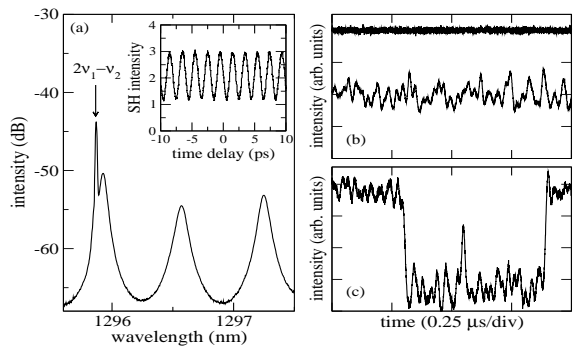}
\caption{\label{auto_corr}  (a) Enlarged spectrum showing the presence of a four-wave mixing sideband.
Inset: Background free intensity autocorrelation measurement showing mode beating at 480 GHz.
(b) Time traces of a primary mode (lower trace) and total device output (upper trace) showing essentially 
constant total output. Not shown is the second primary mode output which is anticorrelated with the first. 
(c) Time trace of the output of a single FP laser mode from the device of Fig. \ref{spectra} (e). At this 
current the plain FP laser is exhibiting mode-hopping behaviour.}
\end{figure}

In Fig. \ref{auto_corr} (a) we have plotted a higher resolution picture of part of the spectrum of Fig. 
\ref{spectra} (d). One can see sideband formation due to four wave mixing (FWM) processes in the cavity. 
FWM is a third order nonlinear process which occurs due to the formation of a dynamic grating in the 
material complex index \cite{shen, park}. The grating is formed through the beating of the primary 
modes at $\nu_{1}$ and $\nu_{2}$ as shown in  Fig. \ref{spectra} (d). The sideband shown appears at a 
frequency of $2\nu_{1} - \nu_{2}$ with the scattering of the primary mode at $\nu_1$ by the grating. 
Within the bandwidth of the inverted semiconductor, the patterned FP cavity naturally provides gain 
and resonant feedback for the FWM sideband, which is slightly detuned from the amplified 
spontaneous emission peak due to the material dispersion and is resonantly enhanced as the two 
primary modes attain equal time averaged intensity. The presence of a large, narrow linewidth signal, 
implies that the two modes are oscillating simultaneously with good phase stability. 

Phase coherence of the primary modes implies an ultrafast intensity modulation of the laser output. 
We measured this mode beating in the cw output of the laser at the difference frequency, $\nu_{2} 
- \nu_{1} \sim 480$ GHz. The result of the background free intensity autocorrelation measurement is 
shown in the inset of Fig. \ref{auto_corr} (a) where the contrast ratio observed is close to the 
theoretical limit of 3:1. In the mode locking regime, the extension of concepts described here to 
create a comb of discrete wavelengths can lead to compact sources of pulsed radiation with THz 
repetition rates. In fact, the polarization associated with the mode beating itself is a source of 
direct THz radiation, generated by the two primary waves in an intracavity process \cite{hoffmann}. 
In an intracavity process a single material system simultaneously acts as the pump medium and 
nonlinear mixing material. Such systems have received considerable interest on account of their 
potential to enhance nonlinear interactions. Approaches to intracavity wavemixing have tended to 
focus on engineering of the semiconductor heterostructure to support distinct optical transitions
and thus multiple lasing wavelengths \cite{capasso, belyanin}. By using the extension of the FP 
cavity concept described here to suppress unwanted longitudinal cavity mode structure, the efficiency 
of such approaches can be further improved. 

We now apply the classical theory of a  homogeneously broadened two mode laser to our device near the 
current of Fig. \ref{spectra} (d). In classical laser theory, simultaneous lasing of two modes is possible 
if the net gain of each mode is positive and the competition due to cross-saturation is sufficiently weakened 
by a large mode spacing and by the spatial hole burning effect associated with the standing waves of the 
FP cavity \cite{sargent}. 

Measurement of a large intensity modulation at the difference frequency of the two modes indicates that the 
device is not exhibiting the mode hopping behaviour characteristic of multiwavelength semiconductor lasers 
with strongly coupled lasing modes \cite{hioe}. We confirmed this by measurements of time traces of the modal 
and total intensity output of the two-color device as shown in Fig. \ref{auto_corr} (b). We observe an 
essentially constant total output and anticorrelated, enhanced intensity noise traces in each of the two primary modes 
due to mode partition (a single modal intensity time trace is shown for clarity) \cite{agrawalpra}. The plain FP 
laser of Fig. \ref{spectra} (e) also shows a constant total intensity, but analysis of individual modal 
intensities reveals complex dynamics including mode hopping behaviour, an example of which is shown in Fig. 
\ref{auto_corr} (c). This figure shows a sequence of spontaneous switching events where the longitudinal mode 
in question switches between an ``on'' state with large intensity to an ``off'' state with an intensity close 
to zero. 

In the two-color device here, the coupling between the primary modes is determined 
by various processes that have different strengths depending on the separation between the modes. These include 
static spectral hole burning, intraband population pulsations and carrier density pulsations. Because the 
characteristic time associated with interband processes is large, the contribution of asymmetric nonlinear 
gain due to the interband carrier density pulsations will be much smaller in the two-color device, 
where the separation between modes is large. If we neglect the asymmetric contribution, weak coupling 
of modes in the two-color device requires $(4/3)\cdot[1 + (\omega_1 - \omega_2)^{2} 
\tau_{\mbox{\tiny in}}^{2}]^{-1} < 1$, where $\omega_{1,2} = 2\pi\nu_{1,2}$ and $\tau_{\mbox{\tiny in}}$ 
is the intraband relaxation time \cite{ogita}. An estimated range of values for $\tau_{\mbox{\tiny in}}$ 
is 100 - 200 fs, which determines a minimum spacing for the two modes of 460-920 GHz. Although this 
estimate is in agreement with the actual modal separation and stability properties observed, a systematic 
study of two-color and other multiwavelength FP lasers will be of interest in order to understand 
separately the roles of the primary mode spacing and the total mode number in determining the stability 
and dynamical properties of this family of devices. 

In conclusion, we have introduced a class of multiwavelength Fabry-P\'erot lasers where the number 
and spacing of the lasing modes is limited only by the bandwidth of the active medium. Measurements 
of simultaneous lasing in a specially designed two-color Fabry-P\'erot cavity geometry with THz
mode spacing were presented. The inverse scattering approach to multiwavelength laser design 
described is likely to open up new avenues for the fundamental studies of semiconductor laser 
stability and dynamics. In addition, the devices can provide interesting and novel solutions to 
many applied problems in optoelectronics and nonlinear optics. 

\textit{Acknowledgments.} This work was supported by Science Foundation Ireland. The authors thank
Guillaume Huyet and John Houlihan for helpful discussions.

\end{document}